\documentclass[useAMS,usenatbib]{mn2e}
\usepackage{amssymb,amsmath}
\usepackage{graphicx}
\usepackage{natbib}
\usepackage{journal_shortcuts}
\newcommand{\bi}{\begin{itemize}}
\newcommand{\ei}{\end{itemize}}
\newcommand{\be}{\begin{equation}}
\newcommand{\ee}{\end{equation}}

\title[Ram pressure histories of cluster galaxies]
{Ram pressure histories of cluster galaxies}

\author[M. Br\"uggen et al.]{M.~Br\"uggen$^{1}$, G.~De Lucia$^{2}$ \\
$^1$ Jacobs University Bremen, P.O. Box 750\,561, 28725 Bremen,
Germany\\
$^2$ Max-Planck-Institut f\"ur Astrophysik, Karl-Schwarzschild-Strasse 1, 85748
Garching, Germany
}

\begin{document}

\date{Accepted. Received; in original form }

\pagerange{\pageref{firstpage}--\pageref{lastpage}} \pubyear{2007}

\maketitle

\label{firstpage}

\begin{abstract}
   Ram pressure stripping can remove significant amounts of gas from
    galaxies that orbit in clusters and massive groups, and thus has a
    large impact on the evolution of cluster galaxies.  In this paper,
    we reconstruct the present-day distribution of ram-pressure, and
    the ram pressure histories of cluster galaxies.  To this aim, we
    combine the Millennium Simulation and an associated semi-analytic
    model of galaxy evolution with analytic models for the gas
    distribution in clusters.  We find that about one quarter of
    galaxies in massive clusters are subject to strong ram-pressures
    that are likely to cause an expedient loss of all gas. Strong
    ram-pressures occur predominantly in the inner core of the
    cluster, where both the gas density and the galaxy velocity are
    higher.  Since their accretion onto a massive system, more than 64
    per cent of galaxies that reside in a cluster today have
    experienced strong ram-pressures of $>10^{-11}$ dyn cm$^{-2}$
    which most likely led to a substantial loss of the gas.
\end{abstract}


\section{Introduction}

In clusters, galaxies can lose some or all of their gas by ram pressure
stripping (RPS) due to their motion through the intracluster medium
(ICM). Both analytical estimates and hydrodynamical simulations show that RPS
can remove a significant amount of gas from galaxies, and can thus explain
observations such as the HI deficiency of cluster disc galaxies (see
e.g.~\citealt{Haynes_Giovanelli_1986,Solanes_etal_2001,cayatte94}), and the
truncated star forming discs in the Virgo cluster
(\citealt{Warmels_1988,koopmann04b}).

RPS only affects the gaseous component of the galaxy so that a distinct
feature of ram pressure stripped galaxies is that their gas discs are distorted
or truncated while their stellar discs appear undisturbed. An increasing number
of observations of ram-pressure stripped galaxies have become available in the
last years (see for example \citealt{Irwin_etal_1987},
\citealt{Veilleux_etal_1999}, \citealt{kenney04}, \citealt{vollmer04a},
\citealt{chung:07}, \citealt{sakelliou:05,sun:07}).

RPS is commonly cited in early work as a possible explanation for the
increased fraction of S0 galaxies in rich clusters relative to the field
(\citealt{Biermann_Tinsley_1975}, \citealt{Butcher_Oemler_1978}). This
explanation was dismissed in the original paper by
\cite{dressler_1980} on the basis of the observation that the relation between
different galaxy populations and local density appears to hold independently of
the cluster richness. Later studies pointed out that additional mechanisms
that lead to a significant redistribution of mass and/or new star formation are
required to explain the entire S0 population of galaxy clusters (see for
example \citealt{moran:07} and references therein). 

The role of RPS in the chemical enrichment of the ICM has also been discussed.
Observational data suggest that the ICM is enriched with metals to
approximately one third of the solar value, suggesting that some fraction of
the metals must have originated from cluster galaxies and since been removed
from them.  Processes responsible for the supply of this enriched gas include
AGN feedback (see e.g. \citealt{roediger:07c} and references therein), galactic
winds driven by supernovae explosions, and ram-pressure stripping
(\citealt{white:91,mori:00,schindler05,domainko:06}). It should be noted,
however, that although RPS certainly affects the metallicity of the ICM, it
may not be the dominant mechanism.  Numerical simulations
by \cite{domainko:06} indicate that RPS can account for only about 10 per cent
of the ICM metal content within a radius of 1.3 Mpc.

The first analytical estimate of RPS dates back to the paper by \citet{gunn72}
who proposed that for galaxies moving face-on through the ICM the success or
failure of RPS can be predicted by comparing the ram pressure with the galactic
gravitational restoring force per unit area.  Later hydrodynamical simulations
of RPS (\citealt{abadi99}, \citealt{quilis00}, \citealt{schulz01},
\citealt{marcolini03,acreman03}, \citealt{roediger:06a},
\citealt{roediger:06b}, \citealt{roediger:07}) suggest that this analytical
estimate fares fairly well as long as the galaxies are not moving close to
edge-on. The ICM-ISM interaction is, however, a complex process influenced by
many parameters. Different aspects have been studied by several groups.
\cite{roediger05} and others have shown that the ICM-ISM interaction is a
multistage process: The most important phases are the instantaneous stripping,
on a time-scale of a few 10 Myr, an intermediate phase, on a time-scale of up
to a few 100 Myr, and a continuous stripping phase that, in principle, could
continue until all gas is lost from the galaxy. 

While numerical simulations and observations indicate that RPS has
important consequences on the amount of gas in cluster galaxies, this
physical process is usually not included in semi-analytic models of
galaxy formation.  The effect of ram-pressure stripping has been
discussed only in a couple of studies using semi-analytic models
(\citealt{okamoto:03, lanzoni:05}), and is shown to produce only mild
variations in galaxy colours and star formation rates.  This happens
because the stripping of the hot gas from galactic haloes
(strangulation) suppresses the star formation so efficiently that the
effect of ram-pressure is only marginal. We note that the studies
mentioned above include ram-pressure stripping based on the analytical
criterion formulated originally in Gunn \& Gott (1972).  In recent
numerical work, \cite{roediger:07} have shown that this formulation
often yields incorrect mass loss rates. Other numerical studies
(e.g. \cite{vollmer01a}) have argued that ram-pressure stripping can
also temporarily enhance star formation. An updated modelling of
ram-pressure stripping that takes into account these results has not
been included yet in semi-analytic models. We plan to address this in
future studies.

For a study of the ram pressure distribution, it is necessary to have
information on the dynamics of galaxies and on the properties of the ICM. The
dynamics of dark matter halos has been studied in a number of papers using
numerical simulations (e.g.  \citealt{benson:05,khochfar:05, diemand:04}).
However, we know of no study on the distribution and history of ram pressures
experienced by galaxies in clusters. If ram-pressure plays some role in
establishing the observed morphological mix in galaxy clusters and/or the
observed radial trends, it is important to quantify the distribution and
history of ram pressures experienced by galaxies that reside in clusters today.
This is the subject of this paper.

\section{Method}

For this study, we rely on the Millenium simulation described in
\cite{springel:05}. This largest dark matter simulation to-date
follows $N=2160^3$ particles of mass $8.5\times 10^8h^{-1}M_{\odot}$
within a comoving box of size $ 500\,h^{-1}$ Mpc a side. The
underlying cosmological model is a $\Lambda$CDM model with
$\Omega_{\rm m}=0.25$, $\Omega_{\rm b}=0.045$,
$\Omega_{\Lambda}=0.75$, $n=1$, $\sigma_8=0.9$ and $h=0.73$, where the
Hubble constant is parametrised as $H_0=100\, h$ km s$^{-1}$Mpc$^{-1}$.
Given its high resolution and large volume, this simulation allows us
to make statistically significant inferences about the ram pressure
histories of galaxies in a representative sample of clusters.

During the simulation, 64 snapshots were saved, together with group catalogues
and their embedded substructures. As explained in \cite{springel:05}, dark
matter haloes are identified using a standard friends-of-friends (FOF)
algorithm with a linking length of 0.2 in units of the mean particle
separation. Each FOF group is then decomposed into a set of disjoint
substructures identified as a locally overdense region in the density field of
the background halo.  The selfbound part of the FOF group itself will then also
appear in the substructure list and represents what we will refer to below as
the main halo. This particular subhalo typically contains 90 per cent of the
mass of the FOF group. The group catalogues were then used to construct
detailed merging history trees of all gravitationally self-bound dark matter
structures. These merger trees form the basic input needed by the semi-analytic
model used in \cite{delucia:07b}.

We extracted the orbital parameters of the galaxies from the public
archive of the Millenium Run data base\footnote{A description of the
publicly available catalogues, and a link to the database can be found
at the following webpage: http://mpa-garching.mpg.de/millennium/}, and
we refer to the original paper for details about the physical
processes that are part of the model.  Our analysis uses only the
orbital parameters of model galaxies and therefore does not rely
explicitely on the details of the semi-analytic model itself. One
important limitation to take into account is that most of the galaxies
in massive clusters are "orphan" galaxies, i.e. galaxies that are not
associated with dark matter substructures. As explained in
\cite{delucia:07b} and in previous papers related to this model,
substructures allow us to trace the motion of the galaxies sitting at
their centres only until tidal truncation and stripping disrupt the
substructures down to the resolution limit of the simulation (which
for the Millennium Simulation is $1.7\times 10^{10} M_{\odot} h^{-1}$)
(e.g. \citealt{delucia:04b, kravtsov:04}).  After this time, the
galaxy is assumed to merge onto the central galaxy of its own halo on
a dynamical friction time-scale, and its position and velocity are
traced by tracking the position and velocity of the most bound
particle of the halo at the last time there was a substructure.
Assuming that the position and velocity of the most bound particle at
the last time the substructure could be identified serve as correct
initial conditions to track the orbit of the galaxy, the ensuing ram
pressure will also be correct.  Recent studies (\citealt{conroy:07})
have argued that a significant fraction of the satellite population
from disrupted subhaloes is unbound and goes to the intra-cluster
light component. A large fraction of galaxies in massive haloes is
represented by orphan galaxies. If, as argued in \cite{conroy:07}, the
model we have used in our study leaves behind an excess of orphan
galaxies, this would affect some of the results presented in this
study. However, the issue regarding orphan galaxies does not seem to
be settled. \cite{wang:06} have shown that orphan galaxies are needed
in order to reproduce the observed correlation function on small
scales. The existence of intra-cluster light suggests that tidal
effects or mergers can unbind some of the stars in the satellite
galaxies. Published results, however, offer little indication of
appropriate recipes for treating this process within semi-analytic
models. Observationally, the total amount of the intra-cluster light
is very difficult to estimate and published estimates vary from less
than 20\% to more than 50\% (see e.g. \citealt{zibetti:05,gonzalez:05}).


As the Millenium run is a dark-matter only simulation, we have to make
assumptions about the distribution of the gas in order to compute the ram
pressure exerted on the cluster galaxies. A first estimate is to approximate
the ICM as isothermal and hydrostatic in a NFW (\citealt{navarro96}) halo whose
profile is given by:

\begin{equation}
\rho_{\rm DM}(r) = {\delta_c \rho_{c0} \over
(r/r_s)(1+r/r_s)^2 } ,
\label{eq:nfwprofile}
\end{equation}
where $\rho_{c0}$ is the critical density of the universe at $z=0$,
and

\begin{eqnarray}
\label{eq:deltac}
\delta_c(M) &\approx& 3\times 10^3 \Omega_0 [1+z_{\rm f}(M)]^3, \\
\label{eq:rs}
r_s(M) &=& {r_{\rm vir}(M) \over c(M)} 
= {1 \over c(M)} \left({3M \over 4\pi\Delta_c\rho_{c0}} \right)^{1/3}.
\end{eqnarray}
In the above expressions, $\Omega_0$ is the density parameter at
$z=0$, $\Delta_c(\Omega_0,\lambda_0)$ is the collapse factor in a
spherical nonlinear model, $z_{\rm f}(M)$ is the average formation
redshift of objects of mass $M$, and $c(M)$ the concentration parameter.

For an isothermal spherical gas cloud with temperature $T_{\rm X}$, the density
distribution $\rho_g$ in hydrostatic equilibrium satisfies the equation:

\begin{equation}
  {kT_{\rm X} \over \mu m_p}{d \ln \rho_g \over dr} = - {G M(r) \over r^2} ,
\label{eq:equilib}
\end{equation}
where $\mu$ and $m_p$ denote the mean molecular weight (we adopt 0.59
below) and the proton mass.  If one neglects the gas and galaxy
contributions to the gravitational mass in the right-hand side, then
the mass enclosed within a radius $r$, can be obtained from
Eq.~(\ref{eq:nfwprofile}) and is given by:  

\begin{equation}
M(r) = 4\pi \delta_c\rho_{c0} r_s^3 m(r/r_s) ,
\label{eq:mhalo}
\end{equation}
where $m(r/r_s)$ is the function $m$ evaluated at $r/r_s$ and $m$ is
given by
\begin{equation}
  m(x) = \ln(1+x)-\frac{x}{1+x} .
 \label{eq:m(x)NFW}
\end{equation}
Equation (\ref{eq:equilib}) can be integrated analytically to give

\begin{eqnarray}
\rho_{\rm g}(r) & = & \rho_{\rm g0}~\exp\left[-{27b \over 2}
\left(1-{\ln(1+r/r_s) \over r/r_s}\right)\right]  \\
& = & \rho_{\rm g0}~\exp(-27b/2)~(1+r/r_s)^{27b/(2r/r_s)} ,
\label{eq:gasprofile}
\end{eqnarray}
with 

\begin{equation}
b\equiv {8\pi G\mu m_p\delta_c(M)\rho_{c0}r_s^2 \over 27kT_{\rm X}}
,
\label{eq:bdef}
\end{equation}
as shown in \cite{makino:98}. The cluster gas temperature $T_{\rm X}$ is
expected to be close to the virial temperature $T_{\rm vir}(M)$ of the dark
matter halo.  In the profile (\ref{eq:nfwprofile}), the latter is in fact
dependent on the radius $r$:

\begin{equation}
kT_{\rm vir}(r) = \gamma {G\mu m_p M(r) \over 3 r},
\label{eq:tvir}
\end{equation}
where $\gamma$ is a fudge factor of order unity which should be determined by
the efficiency of the shock heating of the gas.  \cite{eke:98} adopted
$\gamma=1.5$ as their canonical value in the analysis of X-ray cluster number
counts, and this is also the value we use here. Substituting (\ref{eq:tvir})
into equation (\ref{eq:bdef}), one finds that:

\begin{equation}
b(r) = {2 \over 9\gamma}{r \over r_s} 
\left[\ln\left(1+{r \over r_s}\right)- {r \over r+r_s}\right]^{-1} .
\end{equation}
For an absolute determination of the gas profile, we also need an estimate of
the cluster gas fraction which we chose to be equal to the universal gas mass
fraction $f_{\rm gas}=\Omega_b/\Omega_m= 0.022h^{-2}/0.3=0.14$. There is
evidence that in massive structures such as galaxy clusters this gas fraction
is fairly constant over time (\citealt{allen:07, laroque:06}). The
ram-pressures computed in the following scale simply with this gas
fraction.\\

 As shown by \cite{navarro96}, at large radii the density profile of an
  isothermal gas drops less rapidly than the dark matter (see their Fig.~14).
  This levelling (which is observed outside a radius $\sim 2\times R_{\rm
    vir}$) is not observed in real clusters (\citealt{finoguenov:01b}). The
  model by \cite{makino:98} has been later extended to non-isothermal gas with
  a polytropic equation of state (e.g. \citealt{suto:98}; Ascasibar et al. 2003,
  \citealt{voit:02}).

One alternative approach to reconstruct cluster mass profiles has been
suggested by \cite{komatsu:01}. They present an analytic method to predict gas
density and temperature profiles in dark matter haloes that does not rely on
the isothermal approximation. {In this model, the gas density profile traces
  the dark matter density profile in the outer parts of the haloes (an
  assumption that is also supported by hydrodynamic simulations), and the gas
  obeys a polytropic equation of state. In the inner regions of galaxy
  clusters, gas temperature often increases with radius up to $100-200$~kpc and
  then mildly decreases in the outer regions. The additional assumption that
  the gas temperature has to vary monotonically with density therefore limits
  this model to regions outside the inner $100-200$~kpc. In the model by
  \cite{komatsu:01}, the gas density distribution is given by:}

\begin{equation}
\rho_{\rm gas}(r)= \rho_{\rm gas}(0) y_{\rm gas}(r/r_{\rm s}),
\label{eq:gasprofile}
\end{equation}
where $\rho_{\rm gas}(0)$ is the gas density at $r=0$ and

\begin{equation}
  y^{\gamma-1}_{\rm gas}(x)
  = 1-3\eta^{-1}\left(\frac{\gamma-1}{\gamma}\right)
  \left[\frac{c}{m(c)}\right] [1-\ln(1+x)/x] \ ,
 \label{eq:solution}
\end{equation}
where again we assume a NFW density profile of concentration $c$.
The effective value for $\gamma$ is found to depend weakly on
the concentration parameter according to the following equation:

\begin{equation}
 \label{eq:bestgamma}
  \gamma=1.15 + 0.01\left(c-6.5\right),
\end{equation}
and the parameter $\eta$ is given by

\begin{equation}
 \eta(0)= 0.00676\left(c-6.5\right)^2
          + 0.206 \left(c-6.5\right) + 2.48 .
 \label{eq:eta}
\end{equation}
Equations (\ref{eq:gasprofile})-(\ref{eq:eta}) allow a reconstruction
of the gas profile, and, with the velocity information of the galaxy,
of the ram pressure.

Both models discussed above, the isothermal and the \cite{komatsu:01}
model, rely on the assumption of hydrostatic equilibrium. Numerical
simulations (e.g.  \citealt{ascasibar:03}) show that that this is not
a bad approximation unless they have suffered a major merger in their
recent past. However, there is some indication that the dynamic ICM
can lead to variations in the ram pressure. E.g. it has been suggested
that in the Virgo cluster sloshing motions of the intracluster gas
lead to changes in the ram pressure (\citealt{vangorkom:07}).


\section{Results}

\subsection{Present-day distribution of galactic ram pressures}

In this section, we discuss the distribution of ram-pressures of cluster
  galaxies at the present epoch (i.e. at $z=0$). For this analysis, we have
  selected a number of massive haloes from the Millennium data base, and
  identified all the galaxies from the De Lucia \& Blaizot (2007) semi-analytic
  catalogue that share the same FOF group. For this analysis we have excluded
  the central galaxy of each FOF group (i.e. the brightest cluster galaxies).

Our sample is composed of 20 clusters with masses close to $M=10^{15}M_{\odot}$
($M_{200}$ between $7.3\cdot 10^{14}M_{\odot}$ and $1.2\cdot 10^{15}M_{\odot}$)
and 174 clusters with masses close to $M=10^{14}M_{\odot}$ ($M_{200}$ between
$9.7\cdot 10^{13}M_{\odot}$ and $1.03\cdot 10^{14}M_{\odot}$), yielding a total
of 78,178 and 74,294 galaxies, respectively.

\begin{figure}
\centering\resizebox{\hsize}{!}%
{\includegraphics[angle=0]{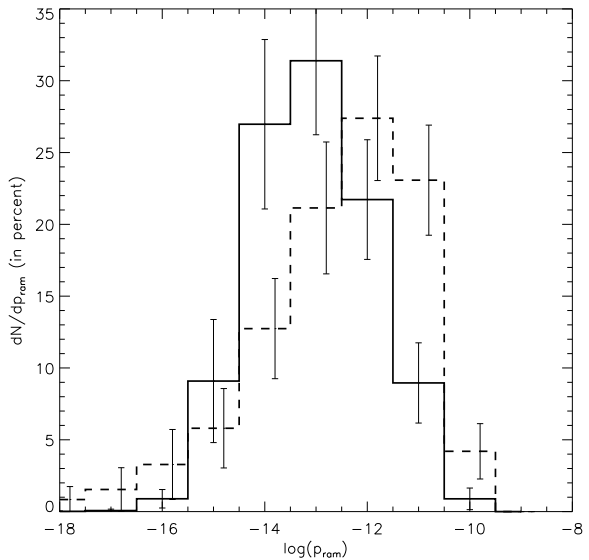}}
\caption{Distribution of galactic ram pressures in isothermal clusters
with virial mass $M=10^{14}M_{\odot}$ (black) and $M=10^{15}M_{\odot}$
(red) at $z=0$. For $M=10^{15}M_{\odot}$, the mean ram pressure is
$10^{-10.8}$ dyn cm$^{-2}$ and the median $10^{-11.8}$ dyn cm$^{-2}$. For
$M=10^{14}M_{\odot}$, the mean ram pressure is $10^{-11.4}$ dyn cm$^{-2}$ and
the median $10^{-12.7}$ dyn cm$^{-2}$. The error bars denote the
cluster-to-cluster scatter.}
\label{fig:dndp}
\end{figure}

\begin{figure}
\centering\resizebox{\hsize}{!}%
{\includegraphics[angle=0]{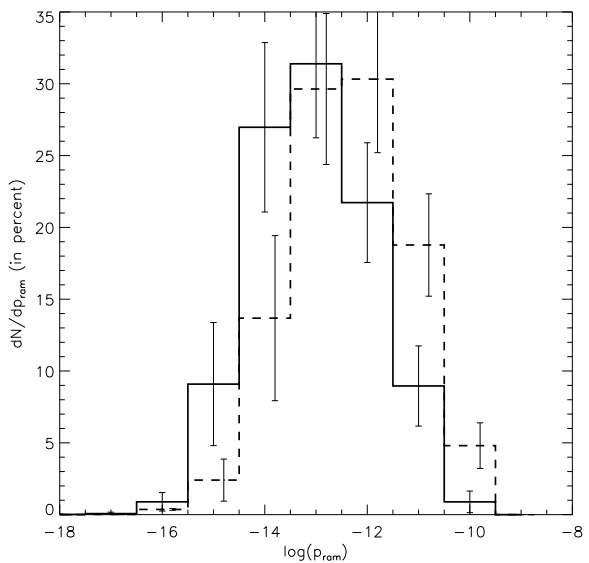}}
\caption{Distribution of galactic ram pressures in clusters with
mass $M=10^{14}M_{\odot}$ (black) and $M=10^{15}M_{\odot}$ (red) at $z=0$
for the model by Komatsu \& Seljak (2001). For $M=10^{15}M_{\odot}$,
the mean ram pressure is $10^{-10.7}$ dyn cm$^{-2}$ and the median ram pressure is
$10^{-11.9}$ dyn cm$^{-2}$. For $M=10^{14}M_{\odot}$, the mean ram pressure is
$10^{-11.3}$ dyn cm$^{-2}$ and the median ram pressure is $10^{-12.6}$ dyn cm$^{-2}$.}
\label{fig:dndp_komatsu}
\end{figure}

Fig.~\ref{fig:dndp} shows the distribution of instantaneous ram pressures
  exerted on galaxies within clusters of masses $M=10^{14}M_{\odot}$ (black
  histogram) and $M=10^{15}M_{\odot}$ (red histogram). For the
$M=10^{14}M_{\odot}$ clusters, the distribution peaks at $\sim 10^{-13}$ dyn
cm$^{-2}$, while for the $M=10^{15}M_{\odot}$ clusters, it peaks at $\sim
10^{-12}$ dyn cm$^{-2}$ (see caption of Fig.~\ref{fig:dndp} for exact values of
the median and the mean).  The shapes of the distributions for both mass ranges
are very similar, with a tail at lower ram pressures and a fairly sharp cutoff
at higher ram pressures. The corresponding plot for the model by
\cite{komatsu:01} is shown in Fig.~\ref{fig:dndp_komatsu}. The
peaks of the two distributions are at similar ram pressure
  values. However, in the \cite{komatsu:01} model the
cut-off at high ram pressures is less sharp, and the distributions appear to be
less skewed. This is because in the isothermal model, the density is
underestimated at large radii. Thus galaxies at large distances from the centre
(which make up the tail of low ram-pressure values that is visible in Fig.~1)
suffer a lower ram pressure in the isothermal model compared to the Komatsu
model. In both Figs.~1 and 2, the solid histograms show the mean obtained
  for all the clusters in each mass bin considered, while the error bars
  indicate the scatter of the distributions.

According to GG72, the mass lost from a galaxy through RPS depends on the
gravitational restoring force per unit area. Clearly, other factors, such as
inclination, gas content, morphological type, star formation rates, etc. also
play a potentially significant, albeit poorly understood, role. The catalogue
compiled by \cite{delucia:07b} contains information that would allow us to
estimate the mass loss due to RPS. However, ram-pressure stripping is not
self-consistently included in the model. Therefore, we focus here on merely
computing the ram-pressures experienced by the galaxies at a given time and
position within the cluster.

In the simulations by \cite{roediger:06a}, ram pressures of
$\sim 10^{-12}$ dyn cm$^{-2}$ were called weak, $\sim 10^{-11}$ dyn
cm$^{-2}$ medium and $\sim 10^{-10}$ dyn cm$^{-2}$ strong. Subject to
strong ram pressure, a typical spiral galaxy as simulated in
\cite{roediger:06a} with mass $\sim 2\cdot 10^{11} M_{\odot}$ loses
all its gas within $\sim 50$ Myrs. Medium ram pressure removes
approximately half of the gas within $\sim 200$ Myrs, and weak ram
pressure removes relatively little gas. These numbers depend of course
on the structure of the gaseous, stellar and dark matter component,
and should just serve for orientation. Figs. \ref{fig:dndp} and
\ref{fig:dndp_komatsu} show that in a $M=10^{15}M_{\odot}$
cluster, approximately 27 per cent of galaxies experience ram pressures of $>
10^{-11}$ dyn cm$^{-2}$ in the isothermal ICM model. In the \cite{komatsu:01}
model, the corresponding fraction is about 24 per cent. In a
$M=10^{14}M_{\odot}$ 
cluster, these numbers are 10 \% for the isothermal model and 9 \% for
the \cite{komatsu:01} model, respectively.

\begin{figure}
\centering\resizebox{\hsize}{!}%
{\includegraphics[angle=0]{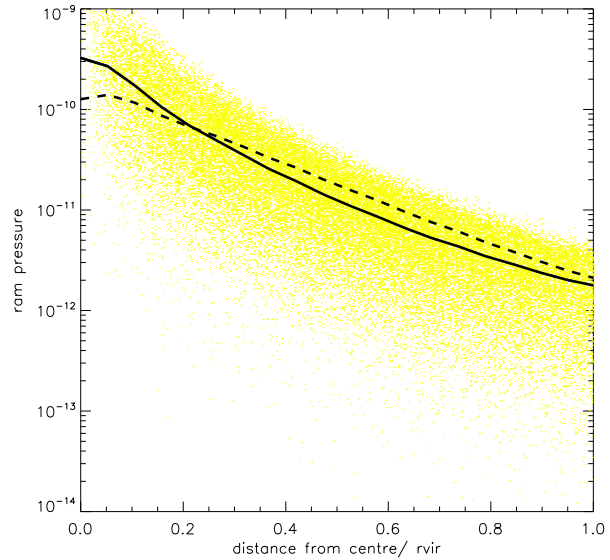}}
\caption{Distribution of galactic ram pressures (in dyn cm$^{-2}$) with radius in a cluster with
mass $M=10^{15}M_{\odot}$ at $z=0$ (model by Komatsu \& Seljak 2001). The thick
solid line denotes the mean of the distribution. The dashed line shows the mean
of the distribution for the isothermal model. The yellow points
indicate the ram pressures of the individual galaxies in the Komatsu
\& Seljak model.}
\label{fig:pr15_k}
\end{figure}

Since ram-pressure depends on the density of the gas and on the
  velocity of the galaxy, it is expected to be stronger closer to the
  centre. In Fig.~\ref{fig:pr15_k}, we plot the ram pressure as a
  function of radius for the Komatsu (2001) model and a
  $10^{15}M_{\odot}$ mass cluster. The distribution shows a sharp
  upper edge which is determined by the escape velocity at this radius
  and the gas density at that position. In the isothermal model, the
  upper envelope can be approximated by:

\begin{equation}
  p_{\rm ram}^{\rm max}(r)\approx \rho_{\rm g} \frac{2GM(r)}{r} .
 \label{eq:plim}
\end{equation}
If we approximate $M(r)$ by Eq.~(\ref{eq:mhalo}) and $\rho_{\rm g}$ by
Eq.~(\ref{eq:gasprofile}), we can rearrange the result to give

\begin{equation}
  p_{\rm ram}^{\rm max}(x)\approx 8\pi G \rho_{\rm
 g0}\delta_c\rho_{c0} r_s^2 \frac{e^{-2B(x)}}{B(x)}(1+x)^{2B(x)/x} ,
 \label{eq:plim2}
\end{equation}
where $x=r/r_s$ and

\begin{equation}
  B(x) = x [\ln(1+x)-x/(1+x)]^{-1} .
\end{equation}
Eq.~(\ref{eq:plim2}) describes the maximum ram pressure pretty well
for $x>0.2$.
Similar, though less simple, expressions can be found for the maximum
ram pressure in the \cite{komatsu:01} model. For the two gas models,
the isothermal and the \cite{komatsu:01} model, the distributions are
somewhat different. While within the virial radius the ram pressures
are very similar, at larger radii, the isothermal model yields lower
ram pressures than the \cite{komatsu:01} model. The latter model
allows the temperature to decrease at larger radii which leads to a
higher density than in the isothermal model. This also explains the
narrower distribution of ram pressures shown in
Fig.~\ref{fig:dndp_komatsu} with fewer galaxies in the low ram
pressure tail of this distribution.

\subsection{Ram pressures histories}

\begin{figure}
\centering\resizebox{\hsize}{!}%
{\includegraphics[angle=0]{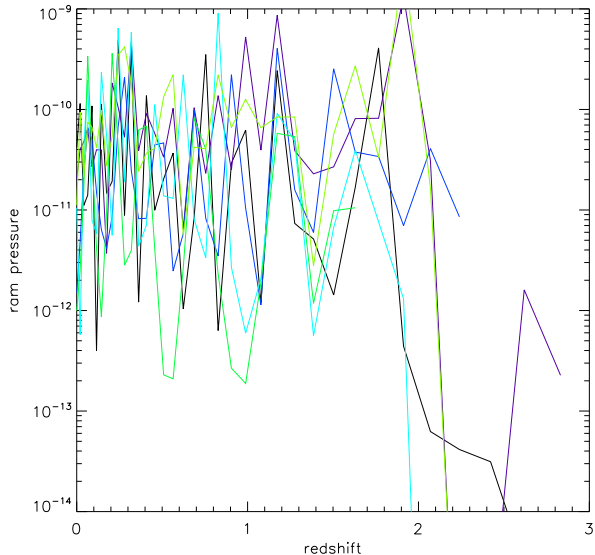}}
\caption{Ram pressure histories of a random sample of galaxies that end up in a
$M=10^{15}M_{\odot}$ cluster.}
\label{fig:history15}
\end{figure}

\begin{figure}
\centering\resizebox{\hsize}{!}%
{\includegraphics[angle=0]{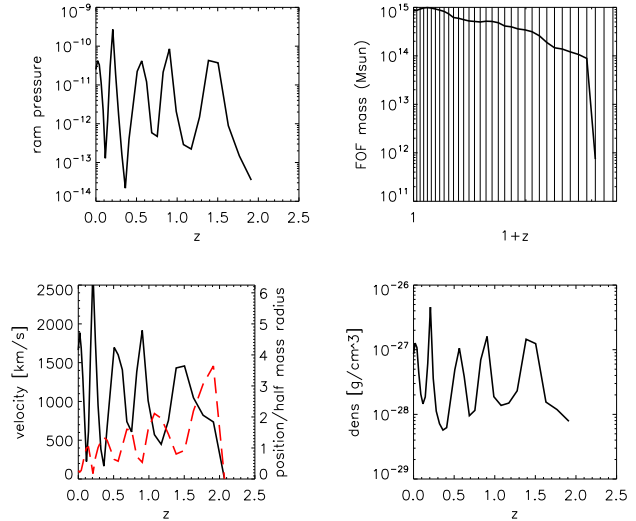}}
\caption{Properties of one selected galaxy versus redshift. Top left: Ram pressure (in dyn cm$^{-2}$). Top right: Mass of the most massive FOF group (i.e. the host cluster or
group) of this galaxy. Vertical lines mark the redshifts of the
simulation output. Bottom left: Velocity (black, solid line) and
distance to cluster centre (red, dashed). Botton right: Ambient ICM
density (using the Komatsu \& Seljak (2001) model.}
\label{fig:multiplot}
\end{figure}

In this section, we analyse the ram-pressure history suffered by
  galaxies that reside in a cluster today. In the following, we
  restrict our analysis to galaxies with with a B-band magnitude
  mag$_B<-19$, where mag$_B$ is given in the De Lucia \& Blaizot
  (2007) catalogue. For each galaxy, we walk its merger tree backwards
  in time by linking it with its most massive progenitor at each
  snapshot until the galaxy becomes a central galaxy of a FOF halo. In
  Fig.~\ref{fig:history15}, we plot a random selection of the ram
  pressure histories of galaxies that end up in $M=10^{15}M_{\odot}$
  clusters. This and the following plots refer to an isothermal ICM
  (the corresponding results for the \cite{komatsu:01} model are very
  similar).  Fig.~\ref{fig:history15} shows that the ram pressures
  fluctuate strongly with time. No strong trend with redshift is
  visible, showing that galaxies underwent phases of strong ram
  pressure even at high redshifts.  In Fig.~\ref{fig:multiplot}, we
  show how the ram pressure (top left panel), the mass of the parent
  FOF group (top right panel), the relative velocity and the distance
  to the cluster centre (bottom left panel), and the ambient ICM
  density (bottom right panel) vary as a function of redshift for a
  randomly selected galaxy that reside in a $M=10^{15}M_{\odot}$
  cluster at the present epoch.  This plot shows that the galaxy
  attains high velocities when it passes close to the cluster core. In
  this region, the ambient ICM density is also highest, such that the
  highest ram pressures values are obtained. We note that the outputs
  of the simulation are not sampled finely enough in time to
  reconstruct the galaxy ram pressure history very accurately.  The
  vertical lines in the top right panel of Fig.~\ref{fig:multiplot}
  show the redshifts of the simulation output. The time between
  snapshots is too large to sample the orbits of the galaxies with
  great precision. In some cases, the ram pressure fluctuates by 1-2
  orders of magnitudes between snapshots and it is difficult to assess
  in each case what happens in between. Consequently, the ram pressure
  histories give conservative bounds on the maximum and minimum ram
  pressures suffered by each galaxy in the course of its life. We also
  note that periods of strong ram pressure are often interspersed with
  periods of weak ram pressure.

\begin{figure}
\centering\resizebox{\hsize}{!}%
{\includegraphics[angle=0]{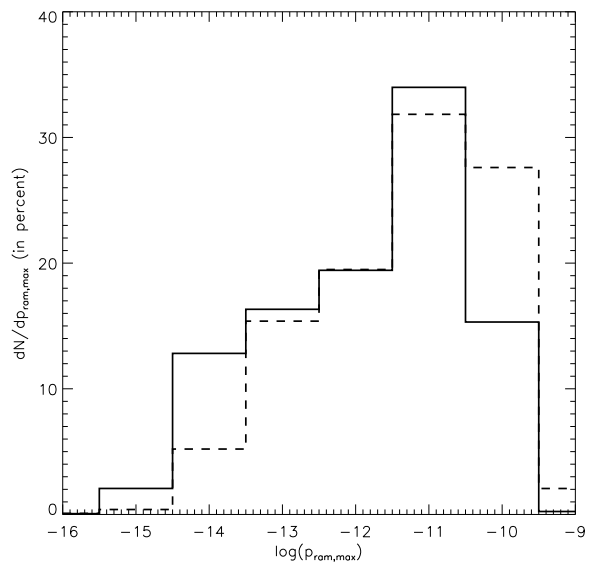}}
\caption{Maximum ram pressure that a galaxy from a
$M=10^{14}M_{\odot}$ (solid line) and $M=10^{15}M_{\odot}$ cluster
(dashed line) or its most massive progenitor has
experienced during its life time.}
\label{fig:maxrp}
\end{figure}

In these periods, the interstellar medium could be accreted back
onto the galaxy or replenished (by e.g. mergers with gas-rich
satellites that have not yet suffered significant stripping). The
median redshift at which a galaxy has most recently experienced
moderate or strong ram pressure is close to 0.1 for all galaxies in
our sample.  Strong
ram-pressure episodes are, however, expected to have a significant
effect on the following evolution of the galaxy. In
Fig.~\ref{fig:maxrp}, we show the maximum ram-pressure that a galaxy
(or its most massive progenitor) has experienced since the time of
accretion. The solid histogram shows results for a
$M=10^{14}M_{\odot}$ cluster, while the dashed histogram refers to a
$M=10^{15}M_{\odot}$ cluster. The figure shows that in a massive
cluster, more than 64 per cent of galaxies have experienced ram
pressures of $>10^{-11}$ dyn cm$^{-2}$, and 32 per cent of galaxies
have had ram pressures greater than $>10^{-10}$ dyn cm$^{-2}$. Ram
pressures of this magnitude most likely strip the galaxy of all its
gas in a short time interval of a few million years. The corresponding
fraction for a $M=10^{14}M_{\odot}$ cluster are lower but still
significant (52 per cent and 11 per cent respectively).  If
ram-pressure stripping is responsible for the morphological
transformation of spiral galaxies infalling onto the cluster from the
field, these numbers indicate that about half of the galaxies residing
today in a $M=10^{14}M_{\odot}$ cluster, and a larger fraction for
more massive clusters, should be gas poor. Ellipticals and lenticulars
make up about $70-80$ per cent of the galaxy population of massive
clusters in the local Universe (see e.g. Dressler 1980). A significant
fraction of these could be therefore entirely explained by
ram-pressure stripping.

\begin{figure}
\centering\resizebox{\hsize}{!}%
{\includegraphics[angle=0]{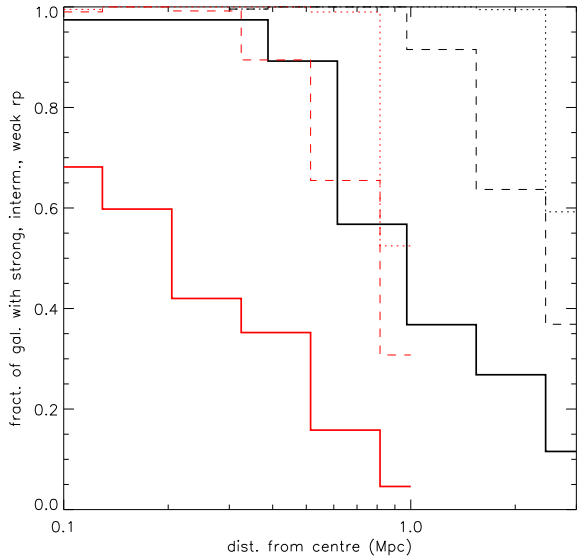}}
\caption{Fraction of galaxies that have suffered strong (solid line),
medium (long-dashed line) and
weak (short-dashed line) ram pressure in the course of their history as a
function of their current ($z=0$) position in the cluster. The black
lines correspond to $M=10^{15}M_{\odot}$ clusters and the red lines
to $M=10^{14}M_{\odot}$
clusters.}
\label{fig:fraction}
\end{figure}

The original work by Dressler (1980) also showed that there is a clear
  trend for an increasing fraction of early type galaxies with decreasing
  distance from the cluster centre. Fig.~\ref{fig:fraction} shows the fraction
  of galaxies that have suffered strong ($> 10^{-10}$ dyn cm$^{-2}$), medium
  ($> 10^{-11}$ dyn cm$^{-2}$) and weak ($> 10^{-12}$ dyn cm$^{-2}$) ram
  pressure as a function of their current ($z=0$) distance from the cluster
  centre. The figure clearly shows that a larger fraction of the galaxies that
  reside in the cluster core have suffered significant ram-pressure. This
  fraction monotonically decreases with distance from the cluster centre, in
  qualitative agreement with the observed trends. In $M=10^{15}M_{\odot}$
  clusters, virtually all galaxies in the inner 300 kpc have suffered strong
  ram pressures after accretion. For $M=10^{14}M_{\odot}$ clusters, this
  fraction decreases to about 2/3. Nearly all galaxies in clusters have
  experienced medium ram pressures and have thus been influenced to some degree
  by ram pressure stripping.

\section{Conclusion}

 In this study we have taken advantage of the Millennium simulation by
  \cite{springel:05} and of the publicly available semi-analytic model by De
  Lucia \& Blaizot (2007) to reconstruct the ram-pressure distribution and
  ram-pressure history of galaxies that reside in clusters at the present
  epoch.  The gas profile in dark matter is described through two analytic
  models which give fairly similar results. We have not included ram-pressure
  stripping self-consistently in the semi-analytic model employed for our
  study. Instead we have simply used the available information about the
  orbital distribution and galaxy merging trees to estimate the importance of
  ram-pressure stripping on galaxies that reside in massive haloes at the
  present epochs.
  
  We find that more than half of the galaxies in a $M=10^{15}M_{\odot}$
  cluster, have experienced ram pressures of $> 10^{-11}$ dyn cm$^{-2}$ after
  their accretion onto a massive system. This fraction is only slightly lower
  for $M=10^{14}M_{\odot}$ clusters, implying that a significant fraction
  of galaxies in clusters at the present epoch suffered substantial gas loss
  due to ram pressure stripping. The fraction of galaxies that suffered
  significant ram-pressure after accretion increases with decreasing distance
  from the cluster centre, in qualitative agreement with the observed increase
  of early-type galaxies.
  
  As expected, strong episodes of ram-pressure occur predominantly in the inner
  core of galaxy clusters, and are restricted to within the virial radius. Our
  results show, however, that virtually all the galaxies in clusters suffered
  weaker episodes of ram pressure, suggesting that this physical process might
  have a significant role in shaping the observed properties of the entire
  cluster galaxy population.
  
  The limited number of simulation outputs does not allow us to reconstruct
  accurately the orbit of the cluster galaxies, and therefore their
  ram-pressure histories. Our result show that ram pressure fluctuates strongly
  so that episodes of strong ram-pressure alternate to episode of weaker
  ram-pressure. During these time intervals, the gaseous reservoir could be
  replenished and new episodes of star formation could occur. Our results
  indicate that ram-pressure stripping must play a significant role in the
  evolution of galaxies residing in massive clusters.  A more self-consistent
  modelling is however required in order to draw more quantitative
  conclusions about the importance and effects of this physical process.

\section*{Acknowledgements}

We thank Simon White for useful comments and suggestions, and Volker Springel
and Gerard Lemson for their help with the Millennium data base.  
MB acknowledges the support by the DFG grant BR 2026/3 within
the Priority Programme ``Witnesses of Cosmic History'' and the
supercomputing grants NIC 2195 and 2256 at the John-Neumann Institut
at the Forschungszentrum J\"ulich.
The Millennium Simulation databases used in this paper and the web application
providing online access to them were constructed as part of the activities of
the German Astrophysical Virtual Observatory.

%
\bibliographystyle{mn2e}
\bibliography{%
/afs/mpa/home/marcus/MYPAPERS/BIBLIOGRAPHY/radio,%
/afs/mpa/home/marcus/MYPAPERS/BIBLIOGRAPHY/metals,%
/afs/mpa/home/marcus/MYPAPERS/BIBLIOGRAPHY/shbib,%
/afs/mpa/home/marcus/MYPAPERS/BIBLIOGRAPHY/marcus,%
/afs/mpa/home/marcus/MYPAPERS/BIBLIOGRAPHY/rphistory,%
/afs/mpa/home/marcus/MYPAPERS/BIBLIOGRAPHY/theory_simulations,%
/afs/mpa/home/marcus/MYPAPERS/BIBLIOGRAPHY/observations_galaxies,%
/afs/mpa/home/marcus/MYPAPERS/BIBLIOGRAPHY/observations_clusters,%
/afs/mpa/home/marcus/MYPAPERS/BIBLIOGRAPHY/galaxy_model,%
/afs/mpa/home/marcus/MYPAPERS/BIBLIOGRAPHY/icm_conditions%
}

\bsp

\label{lastpage}

\end{document}